\documentclass[prl,reprint,superscriptaddress]{revtex4-2}

\usepackage[utf8]{inputenc}
\usepackage{amsmath}
\usepackage{gensymb}
\usepackage{graphicx}

\newcommand{\SrIrO}{Sr$_2$IrO$_4$}
\newcommand{\wn}{cm$^{-1}$}
\begin{document}

\title{Magnetic order, disorder, and excitations under pressure in the Mott insulator \SrIrO}
  
\author{Xiang Li}
\affiliation{Division of Physics, Mathematics, and Astronomy, California Institute of Technology, Pasadena California 91125, USA}
\affiliation{Okinawa Institute of Science and Technology Graduate University, Onna, Okinawa 904-0495, Japan }
\author{S.E. Cooper}
\affiliation{Okinawa Institute of Science and Technology Graduate University, Onna, Okinawa 904-0495, Japan }
\author{A. Krishnadas}
\affiliation{Okinawa Institute of Science and Technology Graduate University, Onna, Okinawa 904-0495, Japan }
\author{A. de la Torre}
\affiliation{Division of Physics, Mathematics, and Astronomy, California Institute of Technology, Pasadena California 91125, USA}
\affiliation{Department of Quantum Matter Physics, University of Geneva, 1211 Geneva 4, Switzerland }
\author{R.S. Perry}
\affiliation{London Centre for Nanotechnology and Department of Physics and Astronomy, University College London, London WC1E 6BT, UK}  
\affiliation{ISIS Facility, Rutherford Appleton Laboratory, Didcot OX11 0QX, UK}
\author{F. Baumberger}
\affiliation{Department of Quantum Matter Physics, University of Geneva, 1211 Geneva 4, Switzerland }
\author{D.M. Silevitch}
\affiliation{Division of Physics, Mathematics, and Astronomy, California Institute of Technology, Pasadena California 91125, USA}
\author{D. Hsieh}
\affiliation{Division of Physics, Mathematics, and Astronomy, California Institute of Technology, Pasadena California 91125, USA}
\author{T.F. Rosenbaum}
\email[Correspondence and requests for materials should be addressed to T.F.R. and Y.F. ]{tfr@caltech.edu and yejun@oist.jp}
\affiliation{Division of Physics, Mathematics, and Astronomy, California Institute of Technology, Pasadena California 91125, USA}
\author{Yejun Feng}
\email[Correspondence and requests for materials should be addressed to T.F.R. and Y.F. ]{tfr@caltech.edu and yejun@oist.jp}
\affiliation{Division of Physics, Mathematics, and Astronomy, California Institute of Technology, Pasadena California 91125, USA}
\affiliation{Okinawa Institute of Science and Technology Graduate University, Onna, Okinawa 904-0495, Japan }

\begin{abstract}
Protected by the interplay of on-site Coulomb interactions and spin-orbit coupling, \SrIrO\ at high pressure is a rare example of a Mott insulator with a paramagnetic ground state. Here, using optical Raman scattering, we measure both the phonon and magnon evolution in \SrIrO\ under pressure, and identify three different magnetically-ordered phases, culminating in a spin-disordered state beyond 18 GPa. A strong first-order structural phase transition drives the magnetic evolution at $\sim$10 GPa with reduced structural anisotropy in the IrO$_6$ cages, leading to increasingly isotropic exchange interactions between the Heisenberg spins and a spin-flip transition to $c$-axis-aligned antiferromagnetic order. In the disordered phase of Heisenberg $J_\mathrm{eff}=1/2$ pseudospins, the spin excitations are quasi-elastic and continuous to 10 meV,  potentially hosting a gapless quantum spin liquid in \SrIrO. 
\end{abstract}
\date{\today}

\maketitle

Mott's treatment of the metal-insulator transition \cite{Mott:1949bv} is a central pillar in the understanding of correlated-electron systems. In $3d$ transition-metal compounds, electron correlations are governed by the on-site Coulomb potential $U$, in relationship with other inter-atomic variables such as the hopping integral, super-exchange interactions, and crystal fields. The intra-atomic spin-orbit coupling $\lambda$ is weak and can be treated in most cases as a perturbation \cite{Mott:1949bv,siCorrelationEffectsIron2009}. In $4d$ and $5d$ transition-metal compounds, the strong $\lambda$ opens an extra dimension in parameter space. The interaction between $U$ and $\lambda$ can lead to new states across insulating, magnetic, and topological phases \cite{witczak-krempaCorrelatedQuantumPhenomena2014}. 

\begin{figure}
    \includegraphics[width=3.25in]{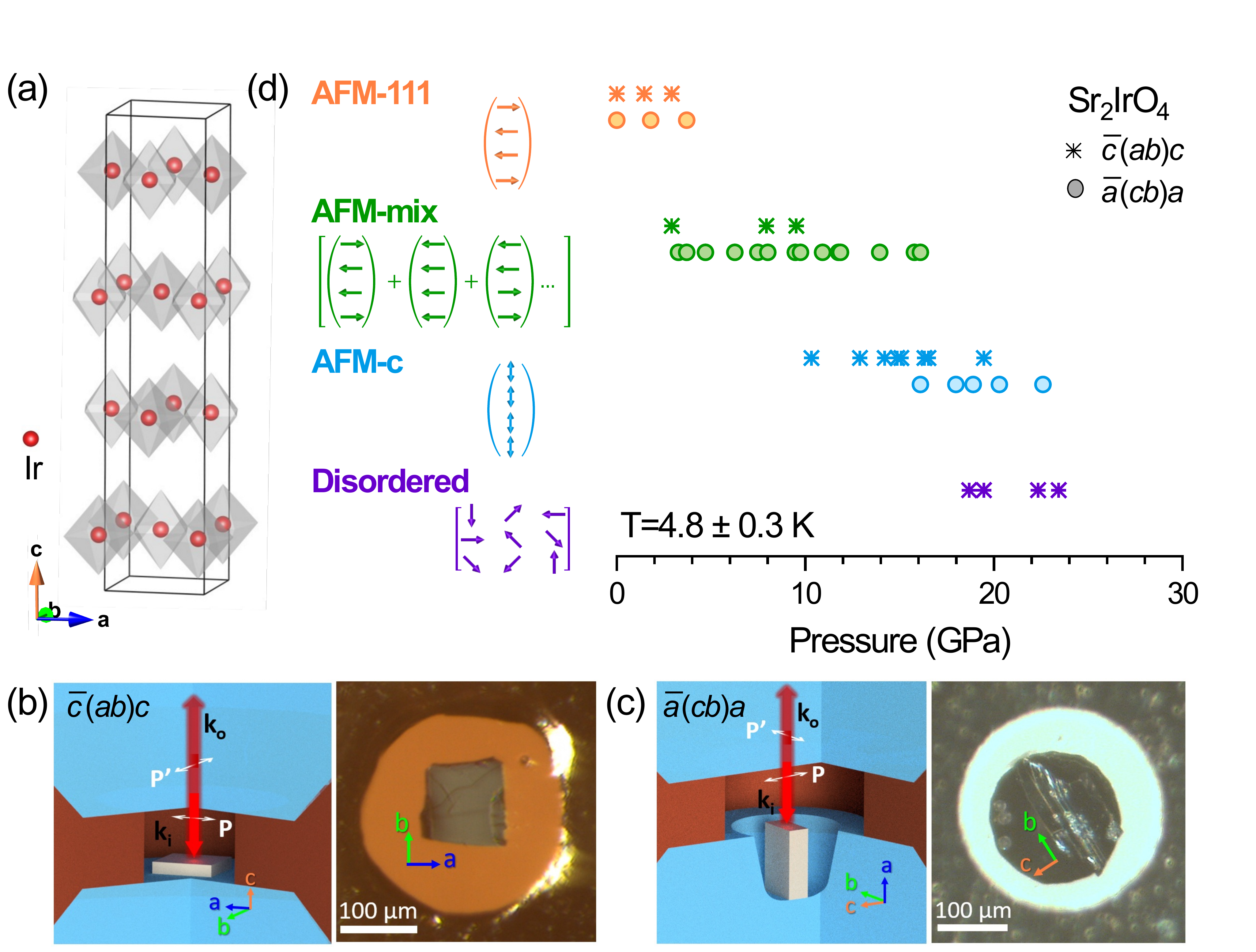}
    \caption{(a) Schematics of \SrIrO\ lattice structure, emphasizing both Ir$^{4+}$ ions (red) and octahedral IrO$_6$ cages (grey). (b, c) Schematics of two  Raman scattering geometries (laser wavevectors $k_\mathrm{i}$, $k_\mathrm{o}$, and polarizations $P$, $P^\prime$) relative to single crystal lattice coordinates $a$, $b$, $c$ inside the high-pressure diamond anvil cell, and photographs of real assemblies.  (d) A summary of magnetic evolution, including schematics of four different spin arrangements under pressure.}
    \label{fig:overview}
\end{figure}

\SrIrO\ exhibits the characteristics of $5d$ spin-orbit-coupling assisted ``Mottness'' \cite{Mott:1949bv,witczak-krempaCorrelatedQuantumPhenomena2014,kimPhaseSensitiveObservationSpinOrbital2009,dimatteoMagneticGroundState2016,phillipsExactTheorySuperconductivity2020}. The insulating state with $J_\mathrm{eff}=1/2$ pseudospins arises from a Coulomb $U$ splitting the half-filled upper $t_{2g}$ band, which is created by the combined effect of crystal field and spin-orbit coupling \cite{witczak-krempaCorrelatedQuantumPhenomena2014,kimPhaseSensitiveObservationSpinOrbital2009,phillipsExactTheorySuperconductivity2020}. The magnetic moments of the Ir$^{4+}$ ions are well protected from many distortions of the local octahedral IrO$_6$ cages (Fig. \ref{fig:overview}a), such as a tetragonal stretch in \SrIrO\ \cite{yeMagneticCrystalStructures2013}, trigonal distortions in both Na$_2$IrO$_3$ \cite{yeDirectEvidenceZigzag2012} and $A_2$Ir$_2$O$_7$ ($A$=Gd, Sm, Eu, Nd, Pr) \cite{wangApproachingQuantumCritical2020}, and a triclinic distortion of the local $x-y-z$ coordinates in Sr$_3$CuIrO$_6$ \cite{liuTestingValidityStrong2012}. Similarly, the measured staggered moment of Ir$^{4+}$ is consistent across systems built on the IrO$_6$ unit, ranging between $0.12-0.37\: \mu_B/\mathrm{Ir}$ \cite{yeMagneticCrystalStructures2013,yeDirectEvidenceZigzag2012,caoAnomalousMagneticTransport2002,guoDirectDeterminationSpin2016,asihMagneticMomentsOrdered2017} in the systems above, smaller than expected for a $J_\mathrm{eff}=1/2$ state \cite{yeMagneticCrystalStructures2013,yeDirectEvidenceZigzag2012}. 

At ambient pressure, \SrIrO\ is antiferromagnetic below $T_\mathrm{N}=240$ K. Within each two-dimensional plane of IrO$_6$ cages, magnetic moments arrange non-collinearly to form a $(1, 1, 0)$ wave vector; they also collectively tilt to form a small ferromagnetic component along either the $a$- or $b$-axis in the square lattice \cite{yeMagneticCrystalStructures2013}. An additional wave vector of $(0, 0, 1)$ is formed by stacking the moments in each plane along the $c$-axis in a “$-++-$” pattern \cite{kimPhaseSensitiveObservationSpinOrbital2009}, leading to an overall $(1, 1, 1)$ wave vector  \cite{yeMagneticCrystalStructures2013}. To understand the underlying physics of the strongly-correlated electronic state, \SrIrO\ has been examined under external tuning parameters such as chemical doping, pressure, and varying IrO$_6$ plane stacking configurations in Sr$_3$Ir$_2$O$_7$ \cite{fujiyamaWeakAntiferromagnetismJeff2012,calderMagneticStructuralChange2012,calderEvolutionCompetingMagnetic2015,Zocco:2014bd,chenPersistentInsulatingState2020,haskelPossibleQuantumParamagnetism2020}. Both Sr$_3$Ir$_2$O$_7$ \cite{fujiyamaWeakAntiferromagnetismJeff2012} and Mn or Ru doped \SrIrO\ \cite{calderMagneticStructuralChange2012,calderEvolutionCompetingMagnetic2015} are antiferromagnets with the Ir moments aligned parallel to the $c$-axis, and insulators even in the paramagnetic phase \cite{fujiyamaWeakAntiferromagnetismJeff2012,calderMagneticStructuralChange2012,calderEvolutionCompetingMagnetic2015}. Under pressure, the insulating phase in \SrIrO\ and Sr$_3$Ir$_2$O$_7$ persists to at least 185 and 104 GPa, respectively \cite{Zocco:2014bd,chenPersistentInsulatingState2020}. By contrast, the antiferromagnetic order in \SrIrO\ is likely suppressed at about 20 GPa \cite{haskelPossibleQuantumParamagnetism2020}. The nature and the number of unique magnetic phases over this pressure range, and a full description of the transitions between them, have not been explored definitively. 
 
 Both the insulating phase and an individual magnetic Ir$^{4+}$ state in the IrO$_6$ cage are protected by the intra-atomic $U$ and  $\lambda$ respectively, and are robust under pressure. By contrast, the antiferromagnetism is not protected and its specific forms are dependent on the detailed balance between various exchange interactions at the inter-atomic level. Here we employ high-pressure, optical Raman scattering \cite{liOpticalRamanMeasurements2020} to probe simultaneously the evolution of the lattice and spin degrees of freedom, and their excitations, in \SrIrO, revealing four unique magnetic phases from ambient pressure to beyond 20 GPa (Fig.\ref{fig:overview}), all connected through first-order phase transitions. At 10 GPa, a change of lattice symmetry from tetragonal to orthorhombic significantly reduces the anisotropy in the IrO$_6$ cage and exchange anisotropy, as confirmed by magnetic Raman scattering. The increasingly Heisenberg-type spins flip to align antiferromagnetically along the $c$-axis. In the high-pressure spin-disordered phase, we measure a broad spin excitation spectrum down to at least 9 \wn\ in the low temperature limit. A high-pressure paramagnetic ground state of $J_\mathrm{eff}=1/2$ Heisenberg pseudospins in the presence of a Mott gap serve as necessary conditions for a quantum spin liquid, and our bounding of spin excitations to energies below 1.1 meV constrains the nature of the paramagnetic state with regards to the potential quantum order. 

\begin{figure}
    \includegraphics[width=3.25in]{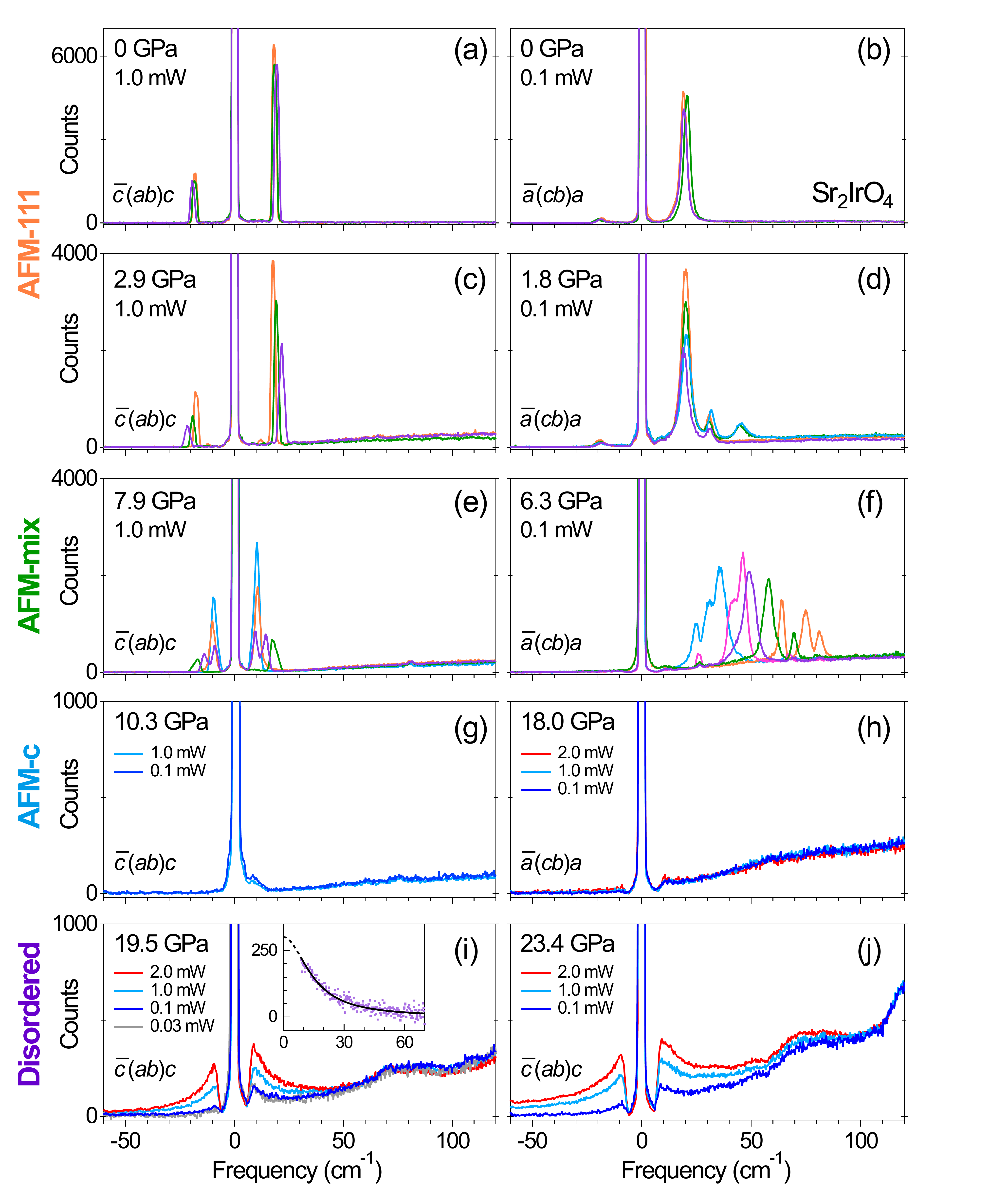}
    \caption{Magnetic Raman spectra, measured in the $\bar{c}(ab)c$ and $\bar{a}(cb)a$ configurations, are shown for (a-d) the AFM-111 phase below 3 GPa,  (e, f) the mixed $c$-axis stacking AFM between 3-10 GPa,  (g, h)  spins aligned parallel to the $c$-axis between 10-20 GPa, and  (i, j) the disordered spin phase above 18.5 GPa. The measured Raman spectra inside the (-9, 9) \wn\ region are dominated by the laser line and the notch filters. All spectra are normalized to a total incident laser exposure of 0.9 J (\textit{e.g.} 0.1 mW over 150 mins). Spectra are presented with measurements either (a-f) at several different spots, or (g-j) at one spot but with different laser powers. The sample temperature is  kept at $4.8 \pm 0.3$ K, but the local temperature within the illuminated $5 \mu m$ focus-spot-size surface region is estimated to be $\sim8$, 17, and 25 K for 0.1, 1 and 2 mW laser powers, respectively, based on ratios of Stokes and anti-Stokes line intensities. (Inset of i) The difference of QES Raman spectra between 0.1 and 2 mW power settings (dots) is fit to a Lorentzian form.
  }
    \label{fig:magnon}
\end{figure}

 To fully access the phonon and magnon modes, our Raman instrument was built to measure inelastic energies down to 9 \wn\ at $4.8$ K and above 20 GPa in two different sample configurations \cite{liOpticalRamanMeasurements2020}. Phonon Raman spectra are measured in the $\bar{c}(CU)c$ and $\bar{a}(CU)a$ configurations, with $\bar{c}/\bar{a}$ and $c/a$ indicating  incident and scattered laser directions in a backscattering geometry to the sample coordinate and parallel to the sample surface normal (Fig. 1). $C$ and $U$ represent circularly polarized (for incident) and unpolarized (for detected) photons, respectively. For magnetic Raman scattering (Fig. \ref{fig:magnon}), the polarizations of the incident and scattered light are kept orthogonal, in $\bar{c}(ab)c$ and $\bar{a}(cb)a$, enhancing the magnon cross section relative to the inelastic charge background. We  frequently checked the $(CU)$ configuration under pressure and noticed no alteration of the observed magnon modes.

\begin{figure}
    \includegraphics[width=3.25in]{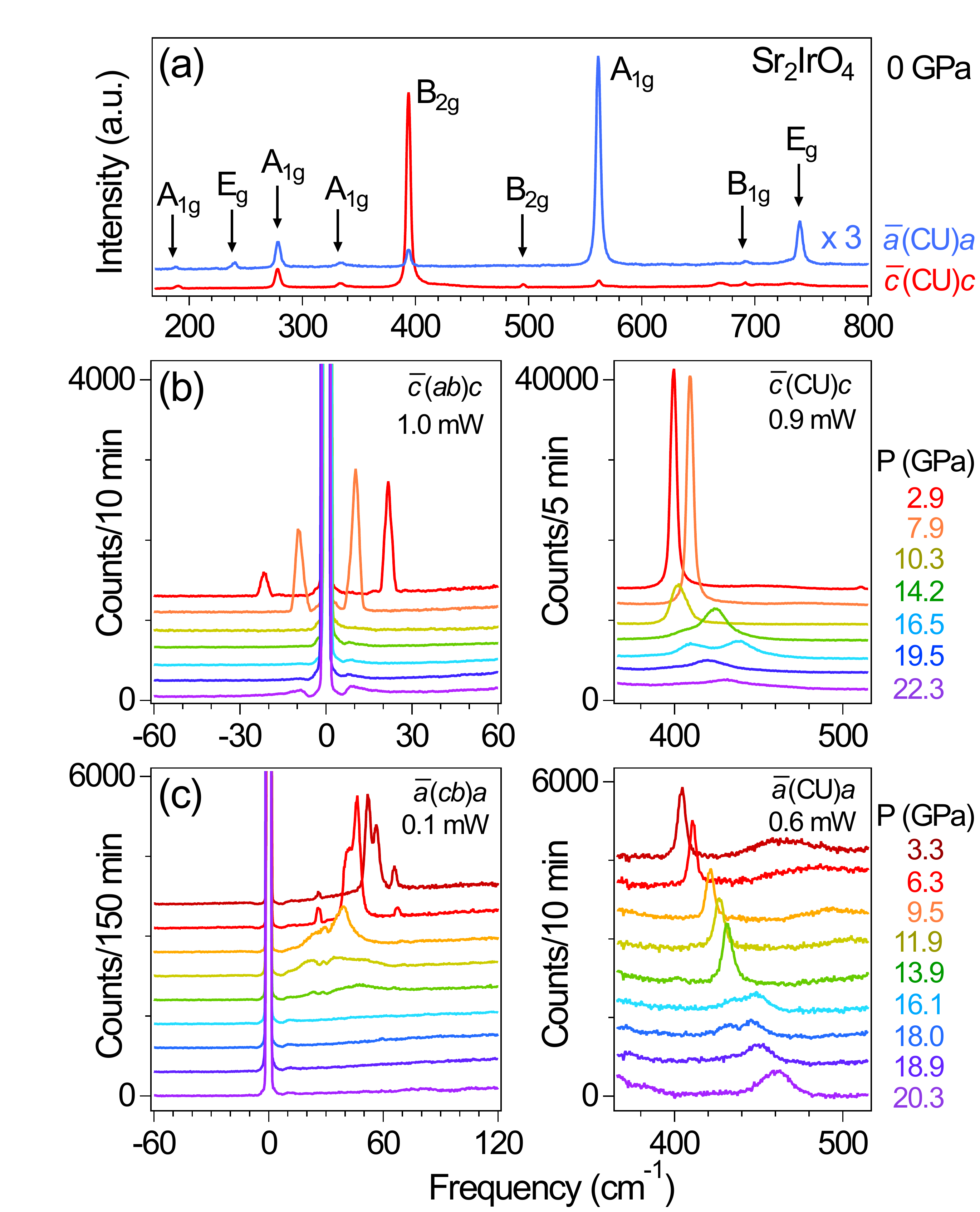}
    \caption{(a) Phonon Raman spectra in $\bar{c}(CU)c$ and $\bar{a}(CU)a$ configurations at ambient pressure and 4.8 K, with  modes marked with arrows and the type. (b, c) Correlated pressure evolution of magnon and phonon behavior.}
    \label{fig:lattice}
\end{figure}

The ambient pressure Raman phonon spectra in both configurations (Fig. \ref{fig:lattice}a) reveal four $A_{1g}$, one $B_{1g}$, and two $B_{2g}$ modes (189, 278, 337, 395, 495, 562, and 692 \wn) that are consistent with Ref. \cite{gretarssonRamanScatteringStudy2017}. We attribute two others (240 and 718 \wn) to $E_g$ modes \cite{gretarssonRamanScatteringStudy2017,cetinCrossoverCoherentIncoherent2012} where a polarization component along the $c$-axis is included (Fig. \ref{fig:lattice}a). Both the $\bar{c}(ab)c$ and the $\bar{a}(cb)a$ scattering configurations identify a single magnon mode at 19-20 \wn\ at ambient pressure (Fig. \ref{fig:magnon}), with otherwise featureless spectra between 9 \wn\ and the $A_{1g}$ mode at 189 \wn. 

From 0 to 24 GPa, we observe distinctive spectroscopic signatures of four magnetic ground states (Fig. \ref{fig:magnon}), all connected through first-order phase transitions with large coexistence regions. The ambient pressure order persists to $\sim 3$ GPa, indicated by  a single magnon  at $\sim$20 \wn\ (Figs. \ref{fig:magnon}a-\ref{fig:magnon}d).

Between 3 to $10^+$ GPa, variations in the magnon spectra emerge, with distinctions between the  $\bar{c}(ab)c$ and $\bar{a}(cb)a$ configurations and spatial variation across the sample surface. The $\bar{c}(ab)c$ magnon spectra typically exhibit two peaks at 10-14 \wn\ and 18-20 \wn, respectively (Fig. \ref{fig:magnon}e), with varying intensity ratios at different sample surface spots, indicating intrinsically different volumes. Previous Raman measurements of \SrIrO\ in the $\bar{c}(ab)c$ configuration with a 0.5 T magnetic field in the $a$-$b$ plane demonstrate a single magnon peak at 10-12 \wn\  \cite{gimIsotropicAnisotropicRegimes2016}, indicative of a $c$-axis stacking pattern of “$++++$” \cite{kimPhaseSensitiveObservationSpinOrbital2009}. While the 20 \wn\ magnon represents the “$-++-$” stacking pattern along the $c$-axis, our observation  points to a coexistence of the “$-++-$” and “$++++$” phases. Furthermore, magnon spectra in the $\bar{a}(cb)a$ configuration demonstrate both the spatial inhomogeneity with different profiles across the sample surface and collective spectral weight  covering the range between 20 and 80 \wn\ (Fig. \ref{fig:magnon}f). While magnon spectra are sensitive signatures of underlying antiferromagnetic order, the large variety, with an overall broad and continuous distribution of magnon energies, indicates that this AFM-mix phase features many types of $c$-axis stacking patterns such as “$-+-+$”, “$-++-$”, and “$++++$” \cite{dimatteoMagneticGroundState2016}, with spatially varying domain composition. The sharp peak profiles of Raman scattering, even at the upper limit of 16.1 GPa \footnote{See Supplementary Materials.}, signify that these stackings are thermodynamic phases with finite correlation lengths along the $c$-axis. Recent resonant x-ray diffraction measurements under pressure \cite{haskelPossibleQuantumParamagnetism2020} suggest that the Ir spins remain confined to the $a$-$b$ plane and order antiferromagnetically within each IrO$_6$ layer. The layers experience a crossover from “$-++-$” to “$++++$” stacking as pressure increases. Our results instead suggest a heterogeneous phase coexistence of many distinct, but energetically close stacking configurations exist for $3-10^+$ GPa. Our magnetic Raman scattering clarifies the ambiguity in Ref. \cite{haskelPossibleQuantumParamagnetism2020} with regard to the extent of both the phase region and the distinctive types of antiferromagnetic order present. 

When the pressure reaches above 10 GPa, the magnon spectra in \SrIrO\ start to disappear in strongly first-order fashion from the Raman sensitive region above 9 \wn\ for both the $\bar{c}(ab)c$ and $\bar{a}(cb)a$ configurations (Figs. \ref{fig:magnon}g-\ref{fig:magnon}h). Despite the absence of low-wavenumber features in this pressure range, the magnetic moments remain ordered in \SrIrO. This is supported by the temperature-independent Raman spectra from 8 to 25 K, set by different incident laser powers of 0.1 to 2.0 mW to locally heat the scattering volume (Figs. \ref{fig:magnon}g-\ref{fig:magnon}h). The measured Raman spectra are identical when normalized by the total flux of incident photons, in sharp contrast to the temperature dependence of spectra in the disordered phase (see below). 

The featureless low-wavenumber Raman spectra (Figs. \ref{fig:magnon}g-\ref{fig:magnon}h) appear simultaneously with changes in the Raman lattice modes, mainly a split of the $B_{2g}$ mode at $\sim$395 \wn\  to a double-peak profile with a significantly-reduced intensity across the pressure phase boundary (Figs. \ref{fig:lattice}b-\ref{fig:lattice}c). This strong correlation is observed both for a single spot on the sample surface at different pressures (Figs. \ref{fig:lattice}b-\ref{fig:lattice}c) and in several different spots across the sample surface at the same pressure \cite{Note1}.  The peak splitting and intensity collapsing behavior of this $B_{2g}$ mode was also observed at 300 K and ~42 GPa in Refs. \cite{chenPersistentInsulatingState2020,samantaFirstorderStructuralTransition2018}, and correlated with a reduction to two-fold symmetry within the $a$-$b$ plane in an orthorhombic structure \cite{chenPersistentInsulatingState2020}. Here, the tetragonal-orthorhombic structural phase boundary moves from  $\sim$42 GPa at 300 K to $\sim$10 GPa at 5 K.

The new set of phonon and magnon Raman spectra suggests a new type of spin order above 10 GPa. Ref. \cite{haskelPossibleQuantumParamagnetism2020} has stated an absence of (1, 0, odd) magnetic reflections to rule out a spin-flip transition to a $c$-axis collinear antiferromagnetic state. However, both raw data and integrated intensities in Figs. 1e and 1g of Ref. \cite{haskelPossibleQuantumParamagnetism2020} demonstrate a significantly-reduced, but still finite amount of  (1, 0, odd) and (1, 0, even) reflections between 10 and 18 GPa. Ref. \cite{haskelPossibleQuantumParamagnetism2020} attributed it to a volumetric suppression of the low-pressure antiferromagnetic “$-++-$” and “$++++$” orders, but left the magnetic state in the majority volume unexplained. We argue that the strong reduction in magnetic diffraction intensities is indicative of the bulk volume experiencing a spin flip transition. 

Previously, a difference of over two orders of magnitude was reported between resonant x-ray magnetic diffraction intensities of \SrIrO\ and Sr$_3$Ir$_2$O$_7$ at ambient pressure \cite{fujiyamaWeakAntiferromagnetismJeff2012}, which was attributed to a reduction of the structure factor between staggered moments aligning either within the $a$-$b$ plane or the $c$-axis. The same argument can be applied to the observed $\sim20\times$ intensity reduction in \SrIrO\ at $\sim$10 GPa \cite{haskelPossibleQuantumParamagnetism2020}. Ref. \cite{haskelPossibleQuantumParamagnetism2020} does not include an x-ray polarization analysis, which is technically feasible under pressure \cite{wangApproachingQuantumCritical2020,Wang:2019jj}, to definitively verify a spin-flip transition \cite{calderEvolutionCompetingMagnetic2015}. The existence of comparable intensities of the (1, 0, odd) and (1, 0, even) reflections between 10-18 GPa in Ref. \cite{haskelPossibleQuantumParamagnetism2020} is similar to the diffraction patterns of $c$-axis-aligned antiferromagnet Sr$_3$Ir$_2$O$_7$ at ambient pressure \cite{fujiyamaWeakAntiferromagnetismJeff2012}, suggesting that the AFM-c phase  have multiple stacking patterns \cite{fujiyamaWeakAntiferromagnetismJeff2012,calderMagneticStructuralChange2012,calderEvolutionCompetingMagnetic2015}, analogous to the AFM-mix phase (Fig. \ref{fig:overview}d).  

Microscopically, the spin-flip process in \SrIrO\ is related to the IrO$_6$ cage distortion. At ambient pressure, the Ir-O bonds in \SrIrO\ are 3.7\% longer along the $c$-axis than those within the $a$-$b$ plane \cite{yeMagneticCrystalStructures2013}. Under pressure, \SrIrO\ has an anisotropic compressibility between the $a$-$b$ plane and the $c$-axis \cite{chenPersistentInsulatingState2020}. While $c$ is less compressible than $a$ and $b$ both below and above the orthorhombic structural transition, at the transition ($\sim$42 GPa and 300 K) it collapses by 9.0\%, whereas $a$ expands by $\sim6$\% and $b$ stays constant \cite{chenPersistentInsulatingState2020,samantaFirstorderStructuralTransition2018}. This results in a substantial reduction in the $c/a$ and $c/b$ ratios and hence a reduced Ir-O bond length anisotropy. In Sr$_3$Ir$_2$O$_7$, the Ir-O bond length anisotropy is only 2.3\% \cite{caoAnomalousMagneticTransport2002}, small enough that the single-magnon mode is not observed in Raman scattering for the $c$-axis aligned spins. The 19 \wn\ one-magnon peak also disappears at ambient pressure in a \SrIrO\ crystal grown in a magnetic field \cite{caoQuestQuantumStates2020}. Study of its antiferromagnetic order could provide a potential comparison, along with optical Raman studies of the $c$-axis spin-order in Mn- and Ru-doped \SrIrO\ \cite{calderMagneticStructuralChange2012,calderEvolutionCompetingMagnetic2015}. 

Above 18.5 GPa, a disordered spin phase appears, characterized by a continuous Raman excitation spectrum (Figs. \ref{fig:magnon}i-\ref{fig:magnon}j) from the lower boundary of instrumental sensitivity at 9 \wn\ to as high as 90 \wn. This inelastic excitation is commonly described as quasi-elastic scattering (QES), and is facilitated by increasing temperature. With clear spatial inhomogeneity across the sample surface,  QES only exists in the high-pressure phase where the antiferromagnetic order in \SrIrO\ is fully suppressed. Using incident laser power to adjust the local sample temperature and the QES intensity, the spectral difference of QES between two temperatures is fit to a Lorentzian form (Fig. \ref{fig:magnon}i inset), indicative of fluctuations  \cite{fleuryDynamicCentralPeaks1983,lemmensCollectiveSingletExcitations2000,wulferdingRamanSpectroscopicDiagnostic2020}. The half-width-at-half-maximum of QES remains large ($\sim$15 \wn) and temperature independent. QES at this energy scale has been observed for magnetic excitations in FePS$_3$ \cite{sekineLightscatteringStudyDynamical1990}, SrCu$_2$(BO$_3$)$_2$ \cite{lemmensCollectiveSingletExcitations2000}, and a variety of quantum spin liquid candidates \cite{wulferdingRamanSpectroscopicDiagnostic2020}, and at thermal structural phase transition in KH$_2$PO$_4$ \cite{kaminowTemperatureDependenceFerroelectric1968}. QES in \SrIrO\ exists only on the high-pressure side of the phase transition, and persists throughout the measured pressure range from 18.6 to 23.4 GPa (Figs. \ref{fig:magnon}i-\ref{fig:magnon}j). Furthermore, the sample temperature of 25 K with 2 mW laser heating is still much lower than the Debye temperature. All these suggest that the QES is not driven by mechanisms specific to a phase transition and softened phonon modes \cite{fleuryDynamicCentralPeaks1983},  but by magnetic fluctuations in the spin-disordered phase. 

Although the lattice structure and phonon Raman modes are different between \SrIrO\ and La$_2$CuO$_4$, the Cu and Ir sublattices are similar, so their magnon spectra are expected to be determined by the same Hamiltonian with different parameter values. The spin Hamiltonian is often expressed as \cite{hayesScatteringLightCrystals1978,gimIsotropicAnisotropicRegimes2016,keimerSoftPhononBehavior1993,gozarMagneticOrderLightly2004}:
$
H = J_{ij}\sum_{\langle ij\rangle}\left[\vec{\mathbf{S}_i}\cdot \vec{\mathbf{S}_j} - \alpha_z S_i^z S_j^z +\boldsymbol{\alpha}_{DM}\cdot(\vec{\mathbf{S}_i}\times\vec{\mathbf{S}_j}  \right] + J\alpha_{\perp}\sum_{\langle ik\rangle}\vec{\mathbf{S}_i}\cdot\vec{\mathbf{S}_k}.
$
The first summation describes interactions between spins $\langle ij\rangle$ within a two-dimensional plane, where the dominant isotropic Heisenberg exchange $J$ is modified by anisotropic Ising spin exchange and Dzyaloshinskii-Moriya (DM) interactions ($\alpha_z J$ and $\boldsymbol{\alpha}_{DM}J$, respectively). 
The last term sums over isotropic Heisenberg interactions between spin pairs $\langle ik\rangle$ on neighboring layers. We neglect additional anisotropic interactions in the Hamiltonian, such as anisotropic interlayer, dipolar, and Jahn-Teller spin-lattice types. 

At ambient pressure, inelastic neutron scattering identified in La$_2$CuO$_4$ two magnon gaps of 1-2 and 3-5 meV at the zone center and an overall energy scale $J\sim150$ meV \cite{keimerSoftPhononBehavior1993,petersTwodimensionalZonecenterSpinwave1988}.  The magnon gaps were attributed to a DM interaction $\boldsymbol{\alpha}_{DM}J$ and an anisotropic exchange interaction $\alpha_z J$, respectively. Optical Raman measurements reveal a single magnon at 17-20 \wn\ in both La$_2$CuO$_4$ and \SrIrO\ at low temperature, zero field, and ambient pressure \cite{gretarssonRamanScatteringStudy2017,gimIsotropicAnisotropicRegimes2016,gozarMagneticOrderLightly2004}. While this mode was identified with the DM anisotropy in La$_2$CuO$_4$ \cite{gozarMagneticOrderLightly2004,petersTwodimensionalZonecenterSpinwave1988}, the large difference in $\lambda$ between Ir and Cu (0.4-0.6 and $\sim$0.01 eV \cite{kimMagneticExcitationSpectra2012}) prevents the 20 \wn\ magnon in \SrIrO\ from being attributed to DM interactions, but instead the anisotropic exchange $\alpha_z J$. With the magnon band gap at the zone center $J{((2+\alpha_z)\alpha_z)}^{1/2}$ \cite{hayesScatteringLightCrystals1978,petersTwodimensionalZonecenterSpinwave1988}, a gap size of 20 \wn\ would lead to $\alpha_z J\sim0.03$ meV in \SrIrO\ at ambient pressure, similar to that in La$_2$CuO$_4$ \cite{petersTwodimensionalZonecenterSpinwave1988}. 

With $T_\mathrm{N}=240$ K and $J\sim60$ meV at ambient pressure \cite{kimMagneticExcitationSpectra2012}, and the absence of QES in ordered phases under pressure, \SrIrO\ demonstrates a low level of spin frustration. Below 10 GPa, the exchange anisotropy $\alpha_z$ is expected to increase as the $c/a$ ratio increases with pressure until the orthorhombic structural transition \cite{chenPersistentInsulatingState2020}. This explains the observed magnon energies reaching as high as 80 \wn (Fig. \ref{fig:magnon}f). However, hydrostatic pressure generally reduces the anisotropy in lattice compressibility \cite{Feng:2012gk}. In the AFM-c state, the reduced IrO$_6$ cage distortion result in a more isotropic local environment and the Ir$^{4+}$ ions approach the full rotational degrees of freedom of Heisenberg spins. The reduced anisotropy $\alpha_z$ diminishes the energy of the zone-center magnon mode, possibly to below the energy sensitivity of Raman scattering.

Raman scattering’s sensitivity to inelastic energies of 1.1 meV is comparable to the energy sensitivity of inelastic neutron scattering measurements \cite{kajimotoStatusNeutronSpectrometers2019,saviciNeutronScatteringEvidence2009}. While magnetic exchange interactions and gapped, discrete dispersion relationships at energies less than 1 meV are well documented \cite{saviciNeutronScatteringEvidence2009}, the continuous spectral weight above 9 \wn\ in Figs. \ref{fig:magnon}i-\ref{fig:magnon}j and the Lorentzian form suggest that the QES in \SrIrO\ extends below 9 \wn\ and the observed spin excitations could be gapless in zero magnetic field. This reflects the reduced exchange anisotropy entering the ordered AFM-c phase continues to higher pressure. A single IrO$_6$ layer in \SrIrO\ could be mapped onto a model of Heisenberg spins on a two-dimensional square lattice, leading to a quantum spin liquid with SU(2) gauge structure and a gapped excitation spectrum \cite{haskelPossibleQuantumParamagnetism2020}, and the broad pressure range of observed QES response can support the necessary stability of a quantum spin liquid. Nevertheless, optical Raman techniques only verify inelastic energy dispersion at $q\approx0$. While our high-pressure Raman scattering provides  a strong upper bound on the gap size and could indicate a gapless scenario, dispositive experimental evidence for a quantum spin liquid awaits additional characterization. For that, neutron or x-ray experiments measuring the complete inelastic spectrum over the entire Brillouin zone are necessary.

\begin{acknowledgments}
Acknowledgments: We thank G. Stenning and D. Nye for help with the instruments in the Materials Characterisation Laboratory at the ISIS Neutron and Muon Source. Y.F. acknowledges support from the Okinawa Institute of Science and Technology Graduate University, with subsidy funding from the Cabinet Office, Government of Japan.  X.L., D.M.S., and T.F.R. acknowledge support from AFOSR Grant No. FA9550-20-1-0263. D.H. acknowledges support from DOE Grant No. DE-SC0010533. F.B. was supported by the Swiss National Science Foundation. A.d.l.T. acknowledges support from the Swiss National Science Foundation through an Early Postdoc Mobility Fellowship (P2GEP2\_165044). 
\end{acknowledgments}



%

\end{document}